\documentclass[superscriptaddress, aps]{revtex4}

\usepackage[english]{babel}
\usepackage{graphicx}
\usepackage{textcomp}
\usepackage{amsmath}
\usepackage{color}
\usepackage{amssymb}
\usepackage{hyperref}
\usepackage{epic}
\usepackage{pict2e}
\begin{document}

\title{Diffusiophoresis at the macroscale}

\author{Cyril Mauger}
\affiliation{LMFA, CNRS and Universit\'e de Lyon, \'Ecole Centrale Lyon, INSA de Lyon and Universit\'e Lyon 1, 69134 \'Ecully CEDEX, France}
\author{Romain Volk}
\affiliation{Laboratoire de Physique, ENS de Lyon, CNRS and Universit\'e de Lyon, 69364 Lyon CEDEX 07, France}
\author{Nathana\"el Machicoane}
\affiliation{Laboratoire de Physique, ENS de Lyon, CNRS and Universit\'e de Lyon, 69364 Lyon CEDEX 07, France}
\author{Micha\"el Bourgoin}
\affiliation{LEGI, CNRS, Universit\'e Joseph Fourier, Grenoble INP, 38041 Grenoble CEDEX 9, France}
\affiliation{Laboratoire de Physique, ENS de Lyon, CNRS and Universit\'e de Lyon, 69364 Lyon CEDEX 07, France}
\author{C\'ecile Cottin-Bizonne}
\affiliation{Institut Lumi\`ere Mati\`ere, CNRS and Universit\'e de Lyon, Universit\'e Lyon 1, F-69622 Villeurbanne CEDEX, France}
\author{Christophe Ybert}
\affiliation{Institut Lumi\`ere Mati\`ere, CNRS and Universit\'e de Lyon, Universit\'e Lyon 1, F-69622 Villeurbanne CEDEX, France}
\author{Florence Raynal}
\affiliation{LMFA, CNRS and Universit\'e de Lyon, \'Ecole Centrale Lyon, INSA de Lyon and Universit\'e Lyon 1, 69134 \'Ecully CEDEX, France}

\begin{abstract}

Diffusiophoresis, a ubiquitous phenomenon that induces particle transport whenever solute concentration gradients are present, was recently observed in the context of microsystems and shown to strongly impact colloidal transport (patterning and mixing) at such scales. 
In the present work, we show experimentally that this nanoscale mechanism can induce changes in the \textit{macroscale mixing} of colloids by chaotic advection. 
Rather than the decay of the standard deviation of concentration, which is a global parameter commonly employed in studies of mixing, we instead use multiscale tools adapted from studies of chaotic flows or intermittent turbulent mixing: concentration spectra and second and fourth moments of the probability density functions of scalar gradients.
Not only can these tools be used in open flows, but they also allow for scale-by-scale analysis. 
Strikingly, diffusiophoresis is shown to affect all scales, although more particularly the small ones, resulting in a change of scalar intermittency and in an unusual scale bridging spanning more than seven orders of magnitude. 
By quantifying the averaged impact of diffusiophoresis on the macroscale mixing, we explain why the effects observed are consistent with the introduction of an effective P\'eclet number.

\end{abstract}

\maketitle

\section{Introduction}

Diffusiophoresis is responsible for transport of large colloidal particles under the action of solutes \cite{bib:Anderson1989,bib:Abecassisetal2009}. 
In the case of electrolyte (salt) concentration gradients, as will be considered in this paper, two mechanisms are involved, both connected to the presence of a nanometric electrical double layer on the surface of the colloid \cite{bib:Anderson1989}: 
the first is purely mechanical and can be explained as a consequence of the existence of gradients of excess of osmotic pressure inside the double layer, while the second is due to electrophoresis of particles in the electric field induced by the difference in mobility of positive and negative salt ions. 
Interestingly, both contributions lead to an additional transport term for the colloids of the same form, proportional to $\nabla \log S$ \cite{bib:Anderson1989}, where $S(\mathbf{x},t)$ is the salt concentration at position $\mathbf{x}$ and time $t$; 
the total contribution is called the diffusiophoretic velocity, denoted $\mathbf{v}_{{\rm dp}}$  (equation \ref{eq:vdp}). 
The equations of motion are thus given by
\begin{eqnarray}
&\frac{\displaystyle \partial S}{\displaystyle \partial t} + \nabla \cdot  [S \mathbf{v}] = D_s\, \nabla^2 S,\label{eq:salt_mixing}\\
&\frac{\displaystyle \partial C}{\displaystyle \partial t} + \nabla \cdot [C (\mathbf{v}+ \mathbf{v}_{\mathrm{dp}})] = D_c\, \nabla^2 C,
\label{eq:col_mixing}\\
&\mathbf{v}_{\mathrm{dp}} = D_{\mathrm{dp}}\; \nabla\log S
\label{eq:vdp},
\end{eqnarray}
where $C(\mathbf{x},t)$ is the colloidal concentration, $\mathbf{v}(\mathbf{x},t)$ is the advecting velocity field, $D_c$ and $D_s$ are the diffusion coefficients of colloid and salt respectively; 
$D_{{\rm dp}}$ is the diffusiophoretic diffusivity.  
This set of equations is valid only if $\mathbf{v}$ is negligibly modified by the movement of the colloids (one way coupling), \textit{i.e.} if the colloidal concentration  is not too large; this is the case here.  
From equation \ref{eq:col_mixing}, it is clear that colloidal concentration is coupled to that of salt via the diffusiophoretic drift velocity, while the salt concentration evolves freely according to equation \ref{eq:salt_mixing}. 

Deseigne \textit{et al.} \cite{bib:Deseigneetal2014} have studied how diffusiophoresis affects chaotic mixing in a micro-mixer (the so-called staggered herringbone mixer \cite{bib:stroocketal02}, $200\,\mu \mathrm{m}$ wide and $115\,\mu \mathrm{m}$ high). 
Using a global characterization ---the normalized standard deviation of concentration, a classical tool in mixing studies--- they observed a diffusiophoretic effect that was interpreted in terms of effective diffusivity (or effective P\'eclet number).
In \cite{bib:Deseigneetal2014}, diffusiophoresis was acting at micron scales and the question remains whether diffusiophoretic effects extend to chaotic mixing at the macroscale: 
will it be able to spread over all length scales or will it remain ineffectively confined at the nano-scale to micro-scales? 
This requires the investigation of possible scale-to-scale coupling: 
while chaotic advection affects all scales of the concentration field from the large scales of the macro-container down to the smallest ones, where diffusion is effective \cite{bib:aref84,bib:ottino89,bib:romkedaretal90,bib:pierrehumbert1994,bib:cerbellietal03,bib:Gouillartetal2006,bib:Lesteretal2013,bib:Gorodetskyietal2015}, what happens when it is combined with diffusiophoresis, a mechanism that originates at the nanoscale? 
In addition to the very existence of the effect, the quantification of its global impact on mixing also needs to be further investigated.

In order to answer these questions, we study diffusiophoresis in a chaotic mixer at the \textit{macroscale}, that is, having dimensions larger than those of microsystems by 2 to 3 orders of magnitudes (up to an overall scale of $5\,\mathrm{cm}$). 
Also, as noted in the abstract, rather than the decay of the standard deviation of concentration, which is a global parameter commonly employed in studies of mixing, we instead apply a set of refined characterizing analyses, using multi-scale tools available from the turbulence community, such as concentration spectra (section \ref{sec:scalar_energy_spectra}), and second and fourth moments of probability density functions (PDF) of scalar gradients (section \ref{sec:concentration_gradients}). 
These more sophisticated tools allow us to perform a scale-by-scale analysis and thus study how all scales of the concentration field are affected by diffusiophoresis. 
Finally, after observing the propagation of diffusiophoretic effects up to the macroscale, we discuss the introduction of an effective P\'eclet number: 
indeed, diffusiophoresis is related to compressible effects through the diffusiophoretic velocity, which is not divergence-free, as shown numerically in \cite{bib:Volketal2014}. 
Thus, it has similarities to the preferential concentration of inertial particles in turbulent flows \cite{bib:maxey1987,bib:shaw2003}.

%%%%%%%%%%%%%%%%%%%%%%%%%%%%%%%%%%%%%%%%%%%%%%%%%%%%%%%%%%%%%%%

\section{Description of the experiment}
\subsection{Experimental set up}
\begin{figure}
\includegraphics{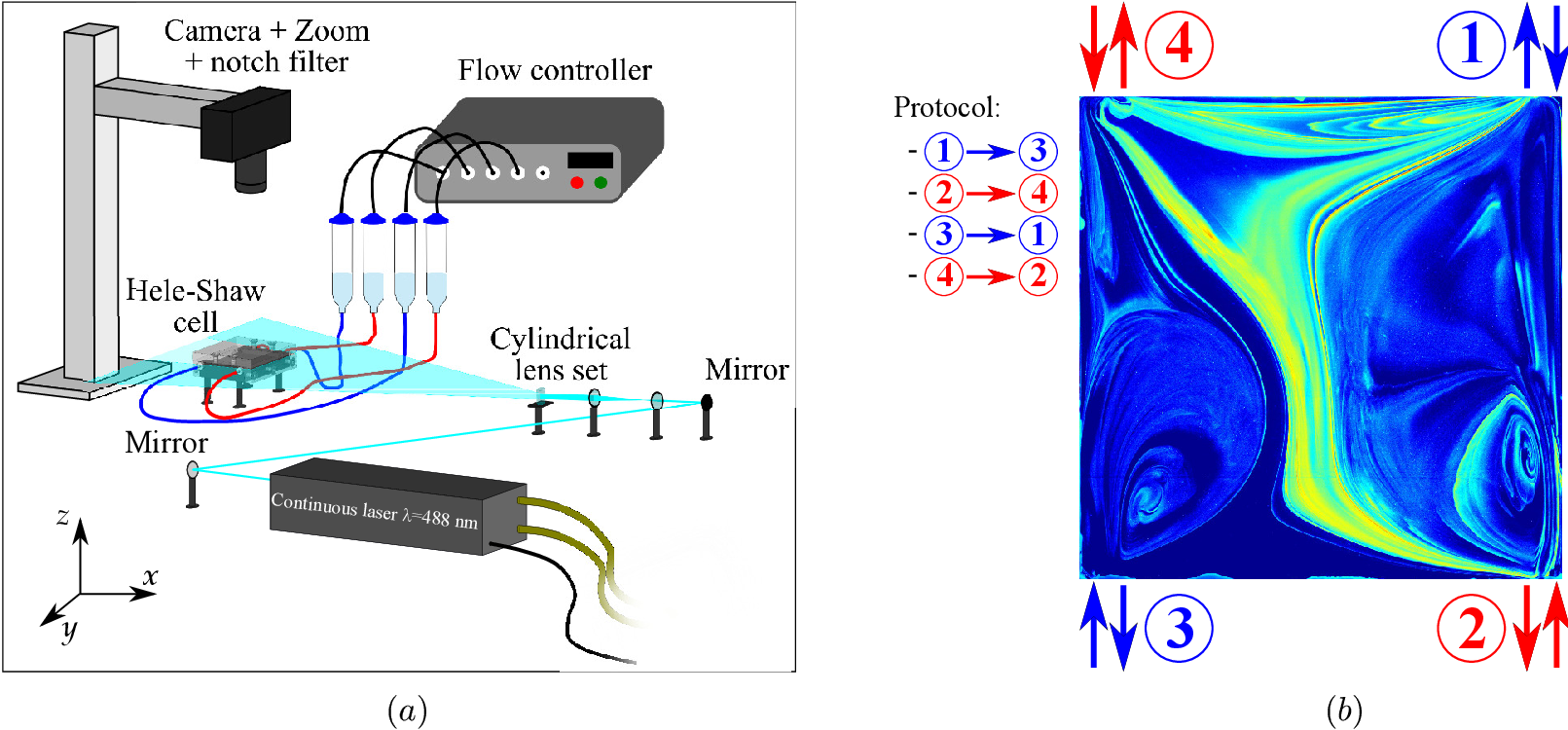}
	\caption{$(a)$ Scheme of the experimental setup; the square Hele-Shaw cell ($L=50\,\mathrm{mm}$) lies horizontally in the $xy$ plane. $(b)$ Time-periodic mixing protocol: \textcircled i $\rightarrow$ \textcircled j indicates that at this step the fluid enters at \textcircled i and exits at \textcircled j during a lapse of time $T/4$, with $T$ the period of the flow-field. The figure displays the instantaneous pattern of a typical concentration field (colloids without salt). A movie showing the observed concentration patterns during the whole  mixing process is also provided as Supplemental Material \cite{ref:movie_colloids}. 
}
\label{fig:setup}
\end{figure}
Mixing takes place in a horizontal, square Hele-Shaw cell of length $L=50\,\mathrm{mm}$ and height $h=4\,\mathrm{mm}$, fitted with four inlets/outlets (figure~\ref{fig:setup}). 
Each inlet/outlet is pressure-driven using a flow controller (Fluigent, MFCS). 
The Hele-Shaw cell is initially filled with water (or salted water, see later). 
At $t=0$, $0.2\;\mathrm{ml}$ of a fluorescent solution (either dye or colloidal suspension) is introduced via inlet~1 into the Hele-Shaw cell using a syringe pump. 
The four inlets/outlets are then pressurized to $100\,\mathrm{mbar}$, and fluid motion is induced by successive pressurization and depressurization of the inlets/outlets:   
a movie showing the mixing process is provided as supplemental material \cite{ref:movie_colloids}. 
Successive deformations of the concentration field are visualised using Planar Laser-Induced Fluorescence (PLIF): 
a continuous laser (Coherent Genesis MX SLM-Series, $\lambda=$~488~$\pm$~3~nm) coupled to a cylindrical lens forms a laser sheet with a typical thickness of the order of the cell height, so that the whole volume of the cell is illuminated. 
The choice of such a thick laser sheet, rather than a thin one localized at the mid-height of
the cell, will be discussed at the end of section \ref{sec:diffusiophoresis}.

The fluorescence signal is recorded with a 14-bit camera (Nikon D700, 4200~px~$\times~2800~\mathrm{pixels}^2$) whose lens (zoom $105\,\mathrm{mm}$) is equipped with a band reject filter (notch $488\,\pm\,12\,\mathrm{nm}$) corresponding to the laser wavelength.
ISO sensitivity is set to the lowest value in order to avoid noise, aperture is set to the highest possible (i.e. $f/3.5$), with a shutter speed of 12.5~ms. 
Image resolution in both horizontal directions ($x$ or $y$) is about $19\, \mu \mathrm{m.px}^{-1}$, while the depth of field is of the order of $1\,\mathrm{mm}$.
Calibration for different fluorescent species and different concentrations showed a linear relationship between the light intensity and the concentration of the species throughout the range studied.

%%%%

\subsection{Flow rate and mixing}
Chaotic advection is produced using the time-periodic protocol illustrated in Figure~\ref{fig:setup}$b$, with four stages of duration $T/4$. 
Efficiency of chaotic mixing in such a Hele-Shaw cell is qualified by the dimensionless pulse volume $\alpha$:
\begin{equation}
\alpha=\frac{qT}{L^2\, h}\, ,
\end{equation}
where $q$ is the flow rate, and $\alpha$ represents the volume of fluid displaced during one period compared to the volume of the chamber \cite{bib:raynaletal2007,bib:beufetal2010}.
For this particular mixing protocol, global chaos (no visible regular region) is obtained for $\alpha\ge1.2$ \cite{bib:raynaletal2004,bib:raynaletal2007}. 
Because large values of $\alpha$ imply rather high flow-rates (hence large Reynolds numbers) or large periods $T$ (hence very long mixing time \cite{bib:raynalgence1995}), we chose to consider the smallest value of interest $\alpha=1.2$. 

In a Hele-Shaw cell, the Reynolds number $\mathrm{Re}_h$ is conventionally based on the height $h$ of the cell, \textit{i.e.}, with typical velocity $q/(hL)$ and kinematic viscosity $\nu$,
\begin{equation}
\mathrm{Re}_h=\frac{q}{L\nu}.
\end{equation}
Note that the Reynolds number inside the pipes connected to the inlets/outlets, 
\begin{equation}
\mathrm{Re}_{pipes}=\frac{4q}{\pi\, d\, \nu}, 
\end{equation}
with $d=1\,\mathrm{mm}$ the diameter of the pipes, is considerably higher. 
Because in the present case  $\mathrm{Re}_{pipes}=64\; \mathrm{Re}_h$, we set $\mathrm{Re}_h=1$ to avoid having too large a Reynolds number in the pipes and hence non-reproducible experiments. 
This corresponds to a flow-rate $q=50\,\mu\mathrm{L}\, \mathrm{s}^{-1}$, and, providing $\alpha=1.2$, we obtain the period $T=120\,\mathrm{s}$.  
Note that with those parameters, the flow is laminar and deterministic, as can also be appreciated in the movie \cite{ref:movie_colloids}. 
As a consequence, the advecting velocity $\mathbf{v}$ in equations \ref{eq:salt_mixing} and 
\ref{eq:col_mixing} is \textit{identical} for all the cases considered here (except for the short initial transient stratification, discussed in appendix \ref{app:strat} for cases with salt). 
Each of the experiments in this article was carried out twice in order to verify that the indicators computed in section \ref{sec:results} were reproducible.

Since we are interested in mixing, the relevant parameter is the P\'eclet number, which measures the relative effect of advection compared to diffusion.
Because in a Hele-Shaw flow chaotic mixing essentially takes place in the horizontal direction \cite{bib:raynaletal2013}, we use the P\'eclet number based on the width $L$ of the cell, 
\begin{equation}
\mathrm{Pe}=\frac{q}{h D},
\end{equation}
where $D$ is the diffusion coefficient of the species considered. 

\begin{table}
\begin{tabular}{|l|c|c|}
\hline
Species &  Diffusion coefficient [$\mathrm{m}^2.\;\mathrm{s}^{-1}$] & P\'eclet number \\
\hline
colloids & $2.\; 10^{-12}$ &  $6.\;10^6$\\
\hline
dextran & $3.6\; 10^{-11}$ & $3.\;10^5$ \\
\hline
fluorescein &  $4.\; 10^{-10}$ &  $3.\;10^4$\\
\hline
salt (LiCl) & $1.4\; 10^{-9}$ & $9.\;10^3$\\
\hline
\end{tabular}
\caption{Species used, diffusion coefficients and corresponding P\'eclet numbers}
\label{table:diffusion}
\end{table}
For this study we used colloids of diameter $200\, \mathrm{nm}$ (FluoSpheres, LifeTechnologies F8811), marked with a yellow-green fluorophore (wavelength $505/515\, \mathrm{nm}$).
In order to characterize the efficiency of mixing as a function of the P\'eclet number (at fixed geometry and flow forcing), other species have also been used, namely fluorescein isothiocyanate (FITC) and fluorescent dextran 70 000 MW (LifeTechnologies D1823). 
For such molecular species diffusiophoresis is not expected to play a role;
they are only used to quantify the deviations induced by diffusiophoresis in the case of colloids with salt. 
The diffusion coefficients and corresponding P\'eclet numbers for all species used in the experiment are available in table \ref{table:diffusion}:  
the variation amplitude of the P\'eclet number is more than two orders of magnitude.

%%%%

\subsection{Diffusiophoresis}
\label{sec:diffusiophoresis}
In order to induce diffusiophoresis, we used a $20\, \mathrm{mM}$ solution of salt (LiCl). 
Indeed, LiCl was shown in microfluidic experiments to have a stronger diffusiophoretic effect than other salts \cite{bib:Abecassisetal2009}: 
for these species (colloid and salt), the diffusiophoretic diffusivity is $D_{\mathrm{dp}}=290\,\mu m^2\,s^{-1}$, and the 
diffusiophoretic motion of the colloids goes from low- to high-salt concentration regions \cite{bib:Abecassisetal2009}.

In the following, we discuss the interplay between mixing and diffusiophoretic drift by considering three different cases:
\begin{itemize}
\item \textit{the reference case}, in which the colloids are injected into pure water;
\item \textit{the salt-in case}, in which the salt is introduced together with the colloids into pure water; in this configuration diffusiophoresis showed hypo-diffusion (delayed mixing) in the staggered herringbone micro-mixer \cite{bib:Deseigneetal2014}.
\item \textit{the salt-out case}, in which the colloids are injected into salted water; in this configuration diffusiophoresis showed hyper-diffusion (enhanced mixing) in the staggered herringbone micro-mixer.
\end{itemize}

%The aim of this study was to prove that diffusiophoresis has visible effects on mixing at the macro-scale, \textit{i.e.} can delay or enhance mixing compared to the reference case. 
%In order to have expertise for comparison with a fixed flow-rate, we also used other fluorescent species, namely fluorescein isothiocyanate (FITC) and dextran 70 000 MW (LifeTechnologies D1823).
%Together with the colloids, this ensures variation amplitude of the P\'eclet number of more than two orders of magnitude (table \ref{table:diffusion}). 

Recall from equations \ref{eq:salt_mixing} and \ref{eq:col_mixing} that, whereas the colloidal concentration  is coupled to that of salt,  the salt concentration $S$ freely evolves during the experiment. 
Thus, the salt is fully mixed for $t\ge L^2\,h\ln(\mathrm{Pe}_s)/(2q) $ \cite{bib:raynalgence97,bib:villermauxetal2008}, with $\mathrm{Pe}_s=q/(h\,D_s)$ the P\'eclet number for salt, that is $t\sim900\,\mathrm{s}$ with our parameters. 
After that time, diffusiophoresis no longer affects the colloids (although the global effect is still visible, \textit{i.e.} mixing enhancement or reduction \cite{bib:Volketal2014}).
In what follows we will restrict attention to times where diffusiophoresis is fully effective. 

Note finally that, because of buoyancy effects, the salt tends to rapidly stratify inside the cell (see appendix \ref{app:strat}). 
Hence, although they have almost the same density as water, the colloids tend to flow from mid-height, where they are injected, towards the bottom of the cell because of vertical diffusiophoresis induced by the salt concentration gradient (appendix \ref{app:strat}). 
This ``settling" of colloids, which is only visible when salt is present and which goes against the effective buoyancy (more salted water in the bottom being denser than colloids), reveals a first macroscopic effect of diffusiophoresis. 
Because it was difficult to follow the colloids over long times using a thin laser sheet (they would eventually disappear below the sheet), and because the flow is quasi-2D, we chose to illuminate the whole cell. 
This kind of height-averaging can result in a loss of signal at small scales, especially in the salt-in case. 
Note that the coupling of the parabolic velocity-profile with diffusion also leads to a vertical homogenization of the concentration field due to Taylor dispersion \cite{bib:Taylor1953}.

%%%%%%%%%%%%%%%%%%%%%%%%%%%%%%%%%%%%%%%%%%%%%%%

\section{Results}
\label{sec:results}
When measuring mixing efficiency, the quantity commonly used is the rate of decay of standard deviation of the concentration $C_{std}(t)=\langle(C-\langle C\rangle)^2\rangle^{1/2}$, or the non-dimensional standard deviation $\sigma(t)=C_{std}(t)/C_{std}(t=0)$ \cite{bib:handbookmix}, where $\langle.\rangle$ stands for the spatial average. 
Indeed, without diffusiophoresis, the rate of decay of $C_{std}(t)$ is related to the presence of high scalar concentration gradients through the equation
\begin{equation}
\frac{d\, C_{std}^2}{dt}=-2 D \langle(\nabla C)^2\rangle.
\label{eq:scalar_dissipation}
\end{equation}
Note that diffusion operates at all scales, but is much more efficient at small scale where the gradients are more intense. 
In the following, as commonly done by fluid mechanicists, the quantity $\frac{1}{2}(C-\langle C\rangle)^2=\frac{1}{2}C_{std}^2(t)$ is referred to as \textit{scalar energy}, by analogy with the kinetic energy.

Above all, chaotic advection involves a large range of scalar scales from the macroscale of the experiment down to the smallest length scale involved, while diffusiophoresis involves a mechanism at the nano-scale. 
Thus such a \textit{global} parameter as $\sigma$ is not enough to explore this typically \textit{multiscale coupled} problem. 
For instance, does diffusiophoresis strongly dissipate scalar energy at a very small scale, or else interact with the flow so as to dissipate more smoothly at all length scales involved? 
In addition, let us note that even for a global characterization, $\sigma$ would not be an appropriate parameter here anyway since the flow is an open flow (marked particles go in and outside the chamber through the inlet/outlets during the periodic mixing protocol, \textit{i.e.} $\langle C\rangle(t)\not=cst$).

In order to investigate the multiscale properties of the concentration field, we used different tools adapted for such a multiscale process: 
\begin{itemize}
\item \textit{the scalar energy spectrum} $E_\theta(k)$ is commonly employed in chaotic advection studies 
\cite{bib:pierrehumbert1994,bib:Williansetal1997,bib:toussaintetal2000,bib:Jullienetal2000,bib:meuniervillermaux2010}; 
it quantifies the scalar energy contained at a given wavenumber $k=2\pi/\ell$, where $\ell$ can be seen as the physical scale at which the scalar energy is calculated, \textit{i.e.} the typical width of a scalar structure; 
it is linked to the \textit{global} scalar energy through the relation 
$\frac{1}{2}C_{std}^2=\int_0^\infty E_\theta(k)\; dk$. 
\item \textit{PDFs of scalar gradients} (more widely encountered in turbulent mixing \cite{bib:Holzer1994,bib:Warhaft2000}, see also \cite{bib:pierrehumbert1994,bib:pierrehumbert2000}); while \textit{global} dissipation of scalar energy is linked to concentration gradients through equation \ref{eq:scalar_dissipation}, such a distribution does allow to investigate whether dissipation occurs mainly with gradients quite close to the mean gradient (as can be seen for instance with a gaussian distribution), or else is related to very intense \textit{local} gradients, in which case we refer to \textit{spatial} intermittency. 
In the present study, each image (corresponding to a given time $t$) allows us to obtain $\mathcal{O} (6.\, 10^6)$ values of the concentration gradient in each direction, further used to compute one PDF.
\end{itemize}

\subsection{Scalar energy spectra}
\label{sec:scalar_energy_spectra}
Instantaneous scalar energy spectra $E_\theta(k)$ are calculated from individual concentration fields at a given time by using the 2D-Fourier-transform $\hat{\theta}(k_x,k_y,t)$ of the reduced concentration field $\theta(x,y,t)=(C(x,y,t)-\langle C(x,y,t)\rangle)/C_{std}(t)$; 
in order to reduce aliasing due to non-periodic boundary conditions, a window-Hanning method was used. 
The 1D isotropic spectrum was then obtained by averaging over each $k=(k_x^2+k_y^2)^{1/2}$. 
\begin{figure}
\includegraphics{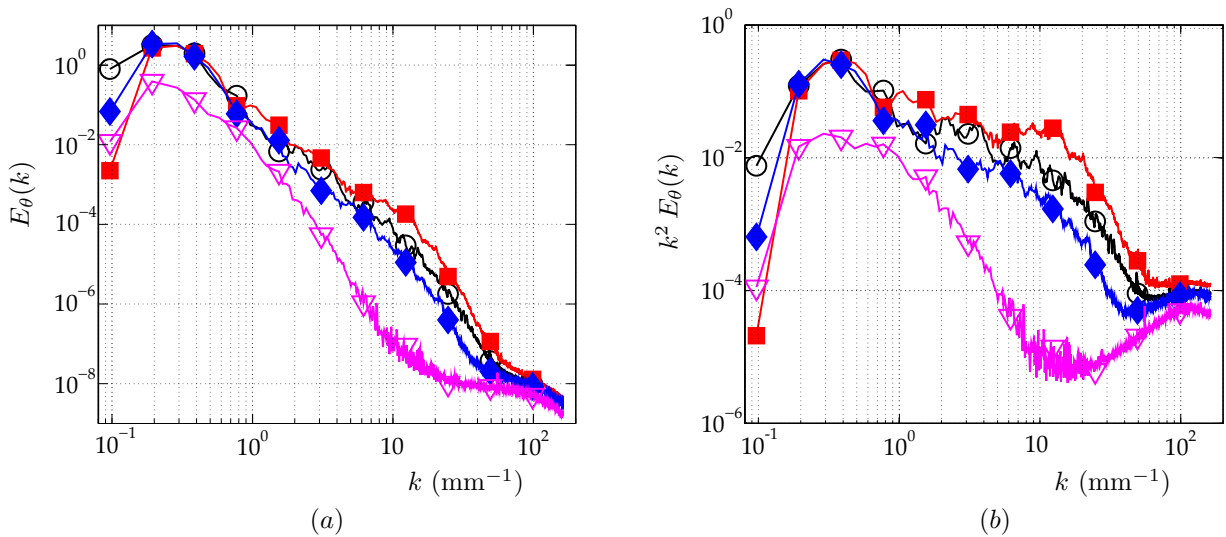}
	\caption{$(a)$ Instantaneous spectra of scalar energy (time $t=160\,\mathrm{s}$); $(b)$ Instantaneous dissipation spectra, same time. 
Open symbols stand for cases without salt. 
$\circ$: reference case (no salt); $\blacksquare$: salt-in; $\blacklozenge$: salt-out; $\triangledown$: fluorescein. The scalar energy spectrum and dissipation spectrum for fluorescein have been divided by 10. 
}\label{fig:inst_spect_conc}
\end{figure}

Figure \ref{fig:inst_spect_conc}$(a)$ shows typical instantaneous scalar energy spectra, plotted on log-log scale, for the three configurations, reference case (without salt), salt-in and salt-out. 
Clearly, the small amount of salt visibly impacts the whole spectrum, although small scalar scales are more affected than large scales (as for diffusion effects).
In the salt-in case (solid squares), the spectrum extends further towards large wavenumbers (small scales) than the reference spectrum.
This kind of behavior would also be observed if considering the concentration spectrum of a species that diffuses less than the colloid we used.
Indeed, since diffusion is directly related to scalar dissipation through equation \ref{eq:scalar_dissipation}, a smaller diffusion coefficient $D$ (therefore a larger P\'eclet number) implies that the final scalar dissipation occurs with larger concentration gradients, \textit{i.e.} at an even smaller lengthscale:
the spectrum would also be shifted towards larger wavenumbers. 
In the salt-out case, the effect is reversed, with a shift towards smaller wavenumbers. 
As a comparison and in order to show the influence of a much smaller P\'eclet number, we have also plotted in the figure the spectrum of fluorescein, although it was divided by 10 for clarity.

The effect is even clearer in figure \ref{fig:inst_spect_conc}$(b)$ when looking at the term $k^2\, E_\theta(k)$, proportional to the scale by scale dissipation budget:
diffusiophoresis obviously affects all lengthscales ranging roughly from the centimeter ($k\ge0.8\, \mathrm{mm^{-1}}$) down to the smallest scales resolved. 
Quite remarkably, this demonstrates that diffusiophoresis can indeed influence mixing processes way beyond its nanometric roots or its micrometric classical influence. 
Combined with chaotic mixing multiscale process, it can spread over more than 7 orders of magnitude in length scales and affect the global system.

However one should note that the previous diagnosis relies on an instantaneous analysis:
while the flow is time periodic, the large scale concentration patterns --and therefore the large scales of the associated spectra-- also vary with time, as can be appreciated on the movie included as supplemental material \cite{ref:movie_colloids}. 
Indeed the effect is not always as pronounced as in figure \ref{fig:inst_spect_conc}; 
at some (rare) moments of the periodic cycle the effect is even reversed, as also found in our numerical simulations \cite{bib:Volketal2014}. 
Because most of the scalar energy is contained in the largest scales (hence in the smallest wavenumbers $k$), and because the large concentration scales vary with time,  
it is not easy to obtain from the spectra a time-averaged parameter that would accurately measure a global effect of salt. 
As observed in the spectra, small scalar scales are more affected by diffusiophoresis: we therefore propose to investigate the scalar gradients, so as to obtain a quantitative comparison that considers a global effect over time.

%%%%	
	
\subsection{Concentration gradients}
\label{sec:concentration_gradients}
\begin{figure}
\includegraphics{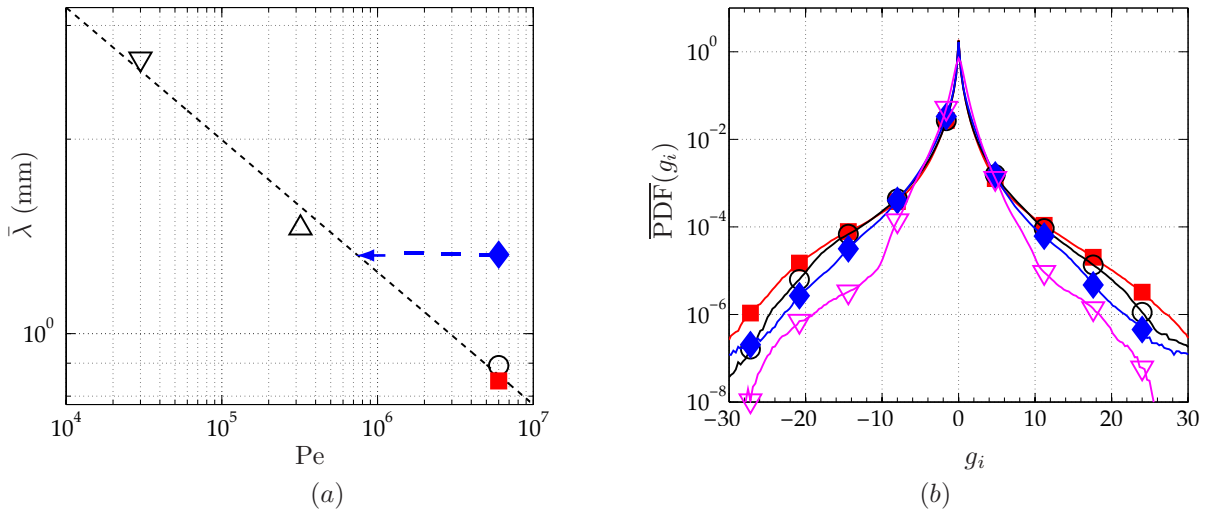}
	\caption{$(a)$ Taylor scale of scalar gradients, $\bar\lambda$, defined as the time-average of $\lambda=2\,C_{std}/\sqrt{G_{x\vert std}^2+G_{y\vert std}^2}$. The arrow indicates the effective P\'eclet number defined as the corresponding reference P\'eclet number that leads to the same value of $\bar\lambda$. $(b)$: time averaged PDF of reduced scalar gradients $g_i$, $\overline{\mathrm{PDF}}(g_i)$. 
Open symbols stand for cases without salt. 
$\circ$: reference case (no salt); $\blacksquare$: salt-in; $\blacklozenge$: salt-out; $\triangledown$: fluorescein. For sake of clarity dextran is omitted in this plot.}
\label{fig:PDF_g}
	\end{figure}

In order to obtain the concentration gradients ${\bf G}= \nabla C$, a given image of the concentration field (corresponding to a given time $t$) is first filtered using a Gaussian kernel to get rid of potential noise: filtering over two pixels ($\approx 40\,\mu m$) is fairly enough to obtain the gradients with great accuracy.  
Then we measure the two components of the concentration gradients, $G_x=\partial C/\partial x$ and $G_y=\partial C/\partial y$ at each point of the image. 
For component $x$ (respectively $y$), we calculate the mean gradient component over the whole image $\langle G_x\rangle$ (respectively $\langle G_y\rangle$),
and also the standard deviation $G_{x\vert std}=\langle(G_x-\langle G_x\rangle)^2\rangle^{1/2}$ (respectively $G_{y\vert std}$).
\begin{figure}
\includegraphics{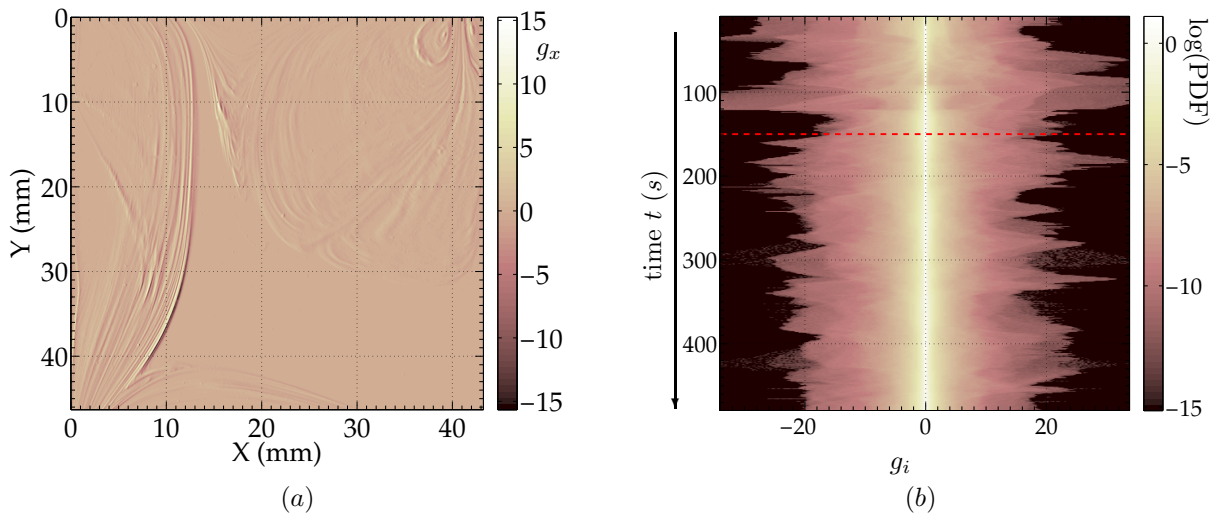}
	\caption{Reference case with colloids (no salt); $(a)$: instantaneous reduced gradient $x$-component, $g_x=(G_x-\langle G_x\rangle)/G_{x\vert std}$, where $G_x=\partial C/\partial x$ (at time $t=293\,\mathrm{s}$). $(b)$: time evolution of $\mathrm{log(PDF)}$ of all reduced gradient components $g_i$, where $g_i$ stands for $g_x$ and $g_y$ ($\textrm{PDF}(g_i)=1/2[\textrm{PDF}(g_x)+\textrm{PDF}(g_y)]$). 
The dotted line at $t=150\,\mathrm{s}$ corresponds to the moment when PDFs become reasonably periodic in time, so that time-averaging is conceivable.}
	\label{fig:pdftime}
	\end{figure}
In the following, we investigate the reduced gradient component $g_i$:
\begin{equation}
g_i=(G_i-\langle G_i\rangle)/G_{i\vert std}\, , 
\end{equation}
where $i$ stands for $x$ and $y$. 
In figure \ref{fig:pdftime}$a$ we plot the reduced gradient $x$-component $g_x$ at a given time ($t=293\,\mathrm{s}$, which corresponds to $2\; 1/4$ periods of the flow-field) in the reference case (no salt). 
Note the very large amplitude range from $-15$ to $15$, indicating that the \textit{spatial} fluctuations of the scalar concentration gradient are not Gaussian (events of large amplitude are more likely to happen than in a Gaussian case, which is commonly referred to as spatial intermittency).
This is reminiscent of the intense and intermittent concentration gradient fronts produced by the mixing process, which are well captured when computing this quantity. 
This results in stretched PDFs of scalar gradient as it will be shown later in figure \ref{fig:PDF_g}$b$. 
While $g_x$ and $g_y$ have equivalent statistics, it is interesting to consider the mean statistics that are even better converged:
figure \ref{fig:pdftime}$b$ shows the PDF of the normalized gradient component $g_i$, $\textrm{PDF}(g_i)=1/2[\textrm{PDF}(g_x)+\textrm{PDF}(g_y)]$, as a function of time (one PDF every second). 
In the experiment, after a transient mixing phase where the initial spot of marked dye begins to spread in the whole domain (roughly one period of the flow-field $T$), the global patterns become almost periodic with time (with period of the flow-field), \textit{i.e.} have a similar shape each period.  
This is also visible in figure \ref{fig:pdftime}$b$ for times $t\ge 150\,\mathrm{s}$ (shown with a dotted line in the figure), where the PDFs have a similar \textit{shape} every period $T=120\,\mathrm{s}$, with abrupt events occurring typically every $T/4$ , \textit{i.e.} associated with a different phase of the periodic protocol (figure \ref{fig:setup}$b$). 
In the sequel we consider time-averaged data, denoted by an over-bar, averaged on the interval of time $150\,\mathrm{s}\le t\le470\,\mathrm{s}$. 
We can now compare cases with or without salt.

For each image (\textit{i.e.} for each time), we define the Taylor length scale associated with concentration gradients as $\lambda=2\,C_{std}/\sqrt{G_{x\vert std}^2+G_{y\vert std}^2}$, and consider its time-average value $\bar\lambda$ (averaged over $150\,\mathrm{s}\le t\le470\,\mathrm{s}$) in figure \ref{fig:PDF_g}$a$. 
When mixing without salt is considered (open symbols, corresponding to cases without any diffusiophoretic effect), $\bar\lambda$ roughly follows a decaying power-law with P\'eclet number. 
In the salt-out case, $\bar\lambda$ is clearly greater than in the reference case. 
We can define an effective P\'eclet number as the corresponding reference P\'eclet number that leads to the same value of $\bar\lambda$ (as suggested by the arrow); 
we obtain a much smaller effective P\'eclet number than for the reference case, $\mathrm{Pe}^{salt-out}_\mathit{eff}\approx8.\,10^5$, that has to be compared to $\mathrm{Pe}\sim 6.\, 10^6$. 
In the salt-in case, the effect is less clear; 
this may be due to the very definition of this quantity, only based on std values of concentration and gradients (second order statistics), which are not as sensitive to the intermittency of the concentration field as are higher order moments. 
In order to check this hypothesis we plot in figure \ref{fig:PDF_g}$b$ the time-average of the instantaneous $\mathrm{PDF}(g_i)$, denoted by $\overline{\mathrm{PDF}}(g_i)$, with or without salt; 
for sake of clarity we omitted the plot for dextran.
When first comparing the cases without salt (open symbols, corresponding to colloids and fluorescein mixing statistics), we recover the usual enhancement of small scale scalar intermittency with increasing $\mathrm{Pe}$ \cite{bib:Holzer1994}: 
the wings of the PDF, plotted on a semi-log scale, are much higher, suggesting that events of large amplitude are more likely to happen. 
When salt is added (closed symbols), once again we recover (with a time-averaged plot rather than the instantaneous ones of figure \ref{fig:inst_spect_conc}) that the salt-out configuration corresponds globally to a smaller effective P\'eclet number. 
In the salt-in case, we observe the effect of a larger P\'eclet number for strong values of gradients, although the plot is hardly distinguishable from the reference case for $-15\le \vert g_i\vert\le15$ (which explains indeed why the two corresponding points are so close in figure \ref{fig:PDF_g}$a$). 
Because the effect of intermittency is more visible on the fourth moment than on the second one, we propose to calculate the flatness of this time-averaged distribution. 
\begin{figure}
\includegraphics{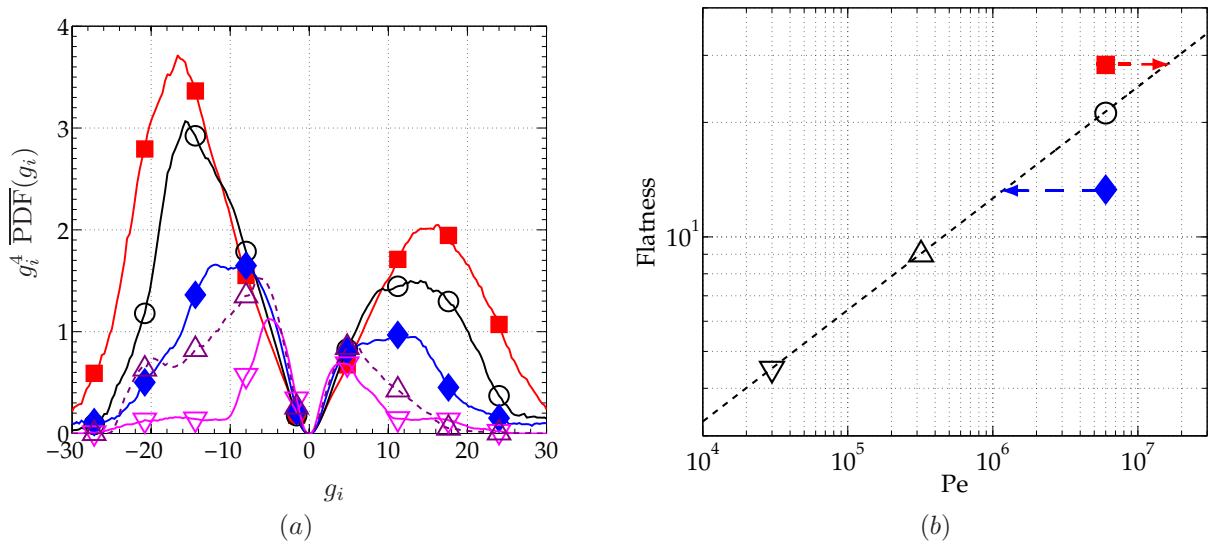}
	\caption{$(a)$: $g_i^4\;\overline{\mathrm{PDF}}(g_i)$,  where $\overline{\mathrm{PDF}}(g_i)$ is the time-averaged PDF of $g_i$ for the cases under study.
 $(b)$: flatness of the time averaged PDF, $F=(\int g_i^4\,\overline{\mathrm{PDF}}(g_i)\;dg_i)/(\int g_i^2\, \overline{\mathrm{PDF}}(g_i)\; dg_i)^2$, with $\|g_i\| \leq 30$). The arrows indicate the effective P\'eclet numbers defined as the corresponding reference P\'eclet numbers that lead to the same flatness.
Open symbols stand for cases without salt. 
$\circ$: reference case (no salt); $\blacksquare$: salt-in; $\blacklozenge$: salt-out; $\triangle$: dextran; $\triangledown$: fluorescein.}
\label{fig:flatness}
\end{figure}	
Indeed, the quantity $g_i^4\;\overline{\mathrm{PDF}}(g_i)$, plotted in figure \ref{fig:flatness}$a$, shows a much pronounced effect in the salt-in case. 
This is even more visible when considering the flatness $F$ of the distribution shown in figure \ref{fig:flatness}$b$: while the flatness in the cases without salt remarkably follows an increasing power-law with the P\'eclet number, the salt-out case rather corresponds to an effective P\'eclet number roughly the same as the one found with the second moment of gradients ($\bar\lambda$), $\mathrm{Pe}^{salt-out}_\mathit{eff}\approx10^6$ ($\mathrm{Pe}^{salt-out}_\mathit{eff}/\mathrm{Pe}\sim 1/6)$, while the salt-in case leads to $\mathrm{Pe}^{salt-in}_\mathit{eff}\approx2.\,10^7$ ($\mathrm{Pe}^{salt-in}_\mathit{eff}/\mathrm{Pe}\sim 3$). 

%%%%

\subsection{Discussion}
\label{sec:discussion}

Overall, our experimental results show that nano-scale diffusiophoresis affects large particles mixing at the macroscale. 
While the results were quantified above using an effective P\'eclet number, it must be kept clear that the underlying mechanism is \textit{not} diffusion. 
Rather, it is related to compressible effects through the diffusiophoretic velocity which is not divergence-free: $\nabla \cdot \mathbf{v}_{\mathrm{dp}}=D_{\mathrm{dp}}\; \nabla^2\log S$ is generally not zero in the presence of salt gradients. 
Thus, although the colloids are transported by the total velocity field $\mathbf{v}+ \mathbf{v}_{\mathrm{dp}}$ (equation \ref{eq:col_mixing}),
the effect is expected to be more complex than a large scale effect through a large ratio of velocity amplitudes $V_{\mathrm{dp}}/V$. 
Indeed, using an order of magnitude estimate, one can prove this ratio to be less than $1\,\%$ here: 
from equation \ref{eq:vdp}, we obtain
\begin{equation}
V_{\rm dp}\sim \frac{D_{\rm dp}}{\ell_s}\ ,
\label{eq:vdp_order}
\end{equation}
where $\ell_s$ is the typical length-scale of salt concentration gradients, which results from a competition of contraction by the chaotic flow-field and diffusion. 
Because the salt is not coupled to the colloids, it obeys \cite{bib:raynalgence97}:
\begin{equation}
\ell_s\sim\frac{L}{\sqrt{\mathrm{Pe}_s}}\ ,
\label{eq:ell_s}
\end{equation}
where ${\rm Pe}_s$ is the P\'eclet number of salt. Finally, from equations \ref{eq:vdp_order} and \ref{eq:ell_s}, we obtain:
\begin{equation}
\frac{V_\mathrm{dp}}{V}\sim\frac{D_\mathrm{dp}}{\sqrt{D_c\, D_s}}\;{\mathrm{Pe}^{-1/2}}\, ;
\label{eq:vdp_rel}
\end{equation}
this order of magnitude is in accordance with what we found numerically \cite{bib:Volketal2014} (with the parameters used for our numerical study we obtain from equation \ref{eq:vdp_rel} that ${V_\mathrm{dp}}/{V}\sim 4.\,10^{-3}$ while we found numerically $6.5\,10^{-3}$ at $\mathrm{Pe}=6.5\,10^4$); 
in our experiment we obtain an even smaller ratio, ${V_\mathrm{dp}}/{V}\sim 2.2\,10^{-3}$. 

Although useful and convenient, the effective P\'eclet approach is only approximate, and is more appropriate in the salt-out configuration where mixing is enhanced. 
In that respect, it is quite remarkable that the effective P\'eclet for the salt-out case is indeed robust against the experimental observable used, either the Taylor scale of scalar gradients or the flatness of the distribution. 
In the salt-in case diffusiophoresis acts \textit{against} diffusion, effectively inducing an ``anti-diffusion'' that strengthens gradients at early times. 
Indeed we have shown in our numerical work \cite{bib:Volketal2014} that a global parameter like the standard deviation of concentration $\sigma$ could increase at small times in the salt-in case, whereas diffusion can only cause $\sigma$ to decrease with time (equation \ref{eq:scalar_dissipation}). 
This is the reason why salt-in characteristics are not easily observed in averaged quantities, and require going to the fourth order moment of the distribution of gradients, rather than the Taylor scale associated to the second moment.
Finally, because diffusiophoresis is related to compressible effects, one could wonder if the use of an effective P\'eclet number is relevant here.  
A first hint can be found in the concentration spectra in figure \ref{fig:inst_spect_conc} (hence in the spatial structures of the concentration field): 
this multiscale approach shows that diffusiophoresis affects all scales of the concentration field, although small scales are even more affected. 
Because the same could be said for diffusion, the effect of diffusiophoresis on the concentration field has some similarities with diffusive effects. 
Another hint derives directly from equation \ref{eq:vdp_rel}: the relative transport by diffusiophoresis compared to that by the velocity field decreases with the P\'eclet number. 
This is also true for diffusion compared to advection from the very definition of the P\'eclet number!
This provides a clue as to why it is useful and meaningful to introduce an effective P\'eclet number when considering the long time effects, and to try to quantify the combined effects of diffusiophoresis and diffusion with that effective approach. 

\begin{table}
\begin{tabular}{|l|c|c|c|c|}
\hline
\null &  $\mathrm{Pe}$ & $D_\mathrm{dp}/\sqrt{D_c\, D_s}$ & $\mathrm{Pe}^{salt-out}_\mathit{eff}/\mathrm{Pe}$ & $\mathrm{Pe}^{salt-in}_\mathit{eff}/\mathrm{Pe}$\\
\hline
Sine-flow (numerical) & $10^4$ & $1$ & $1/2$ & $3/2$\\
\hline
Herringbone mixer (experimental) & $9.\, 10^4$ & $5.56$ & $1/40$ & $20$ \\
\hline
Present experiment & $6.\, 10^6$ & $5.56$ & $1/6$ & $3$ \\
\hline
\end{tabular}
\caption{Three cases where diffusiophoresis is combined to chaotic advection; for the numerical case and the micro-mixer, we used a P\'eclet number in the middle-range of those investigated.}
\label{table:comparison}
\end{table}
In the following we collect results obtained with diffusiophoresis in chaotic advection with different velocity fields: the herringbone micro-mixer \cite{bib:Deseigneetal2014}, the sine-flow \cite{bib:Volketal2014} and the present flow.\\
Although the flow-field in the herringbone micro-mixer is stationary and 3-dimensional, it may however be compared favorably with what can be expected in a two-dimensional time-periodic flow: 
indeed, Stroock \& McGraw \cite{bib:StroockMcGraw2004} proposed an analytical model in which the cross-section of the channel is treated as a lid-driven cavity flow; 
they showed that this model was able to reproduce the advection patterns that were observed experimentally in their flow, whose dimensions are about the same as in Deseigne \textit{et al.} (roughly $200\,\mu \mathrm{m}$ wide, $100\,\mu \mathrm{m}$ high). 
Here, because of the spatial periodicity in the axial direction, the corresponding coordinate plays the role of time. 
Correspondingly, the P\'eclet number in the micro-mixer has to be based on the cross-sectional velocity rather than on the axial velocity. 
With their model, Stroock \& McGraw could also estimate the magnitude of the velocity $u_{\mathrm{cross}}$ in the cross-sectional flow relative to the axial velocity $U$: 
taking $u_{\mathrm{cross}}\sim 0.1\, U$, with a channel width $w=200\, \mu \mathrm{m}$ and $U=8.6\, \mathrm{mm/s}$, we obtain a colloidal P\'eclet number $\mathrm{Pe}\sim 9.\, 10^4$.\\
All the results are summarized in table \ref{table:comparison}. In order to compare numerical and experimental results, we introduced the diffusiophoretic coefficient $D_{\mathrm{dp}}$ (equal to $290\,\mu m^2\,s^{-1}$ in both experiments) using a dimensionless parameter; 
because of equation \ref{eq:vdp_rel} , we chose to compare $D_\mathrm{dp}/\sqrt{D_c\, D_s}$.\\
It is not easy to compare those three cases: not only are the P\'eclet numbers different, but also the diffusiophoretic coefficient is higher in the experiments. Note also that the present flow is an open flow ($\langle C \rangle(t)\not=cst$), while the others are not: for the micro-mixer, $\langle C\rangle(t)=cst$ in all planes perpendicular to the axial direction, and the sine-flow uses periodic boundary conditions. 
However, in all cases, the effect is more important in the salt-out than in the salt-in case. 
 Moreover, for the two experiments where the same colloids and salts were used, we obtain quite a remarkable result, \textit{i.e.} $\mathrm{Pe}^2\simeq2\, \mathrm{Pe}^{salt-out}_\mathit{eff}\, \mathrm{Pe}^{salt-in}_\mathit{eff}$.

%Since the colloids are transported by the total velocity field $\mathbf{v}+ \mathbf{v}_{\mathrm{dp}}$ (equation \ref{eq:col_mixing}), and since the relative effect of diffusiophoresis decreases with the P\'eclet number (equation \ref{eq:vdp_rel}), this could also explain part of the shift (the relative effect is 8 times larger in the micro-mixer because of the difference in P\'eclet number). 
%Note however that  equation \ref{eq:vdp_rel} can not be used to measure directly the difference in effective P\'eclet numbers: 
%indeed, the effective P\'eclet number does not result from diffusiophoresis effects only, but also from diffusion effects, whose relative strength compared to transport by the velocity field also decreases with the P\'eclet number from its very definition. 

%%%%%%%%%%%%%%%%%%%%%%%%%%%%%%%%%%%%%%%%%%%%%%%%%%%%%%%%%%%%%%%%%%%%%%%%%%%%%%%%%%%%%%%%%%

\section{Summary}

In this article we have studied experimentally the effects of diffusiophoresis on chaotic mixing of colloidal particles in a Hele-shaw cell at the macroscale. 
We have compared three configurations, one without salt (reference), one with salt with the colloids (salt-in), and a third one where the salt is in the buffer (salt-out). 
Rather than the decay of standard deviation of concentration, we have used different multiscale tools like concentration spectra, second and fourth moments of the PDFs of scalar gradients, that allow for a scale-by-scale analysis; 
those tools are also available in open flows, when marked particles can go in and out the domain under study. 

Using scalar spectra, we have shown qualitatively that diffusiophoresis affects all scalar scales. 
This demonstrates that this mechanism at the nano-scale has an effect at the centimetric scale, \textit{i.e.} 7 orders of magnitude larger. 
Because the smallest scalar scales are more affected, this results in a change of spatial intermittency of the scalar field: 
using second and fourth moments of the PDFs of scalar gradients, we have been able to quantify globally the impact of diffusiophoresis on mixing at the macroscale.  
Although diffusiophoresis is clearly induced by compressibility effects, we have explained how the combined effects of diffusiophoresis and diffusion are consistent \textit{when averaging in time} with the introduction of an effective P\'eclet number: 
the salt-in configuration corresponds to a larger effective P\'eclet number than the reference case, and the opposite for the salt-out configuration. 
Because this results from a time-averaged study, and not from an instantaneous diagnostic, this demonstrates that diffusiophoresis, a mechanism which originates at the nanoscale, has a quantitative effect on mixing at the macroscale.

%\vskip1cm

%%%%%%%%%%%%%%%%%%%%%%%%%%%%%%%%%%%%%%%%%%%%%%%%%%%%%%%%%%%%%%%%%%%%
 
\acknowledgments 
We are very grateful to Jean-Pierre Hulin and Laurent Talon for really helpful discussions on gravity currents.\\
This collaborative work was supported by the LABEX iMUST (ANR-10-LABX-0064) of Universit\'e de Lyon,
within the program ``Investissements d'Avenir'' (ANR-11-IDEX-0007) operated by the French National Research
Agency (ANR).

%%%%%%%%%%%%%%%%%%%%%%%%%%%%%%%%%%%%%%%%%%%%%%%
\appendix

\section{Absence of gravity currents, stratification of salt and associated migration of colloids}
\label{app:strat}

It may be thought that given the density difference between pure and salted water, we could observe gravity currents inside the Hele-shaw cell at the first stages of the time-periodic flow (before salt begins to mix due to chaotic advection); the arguments given in this appendix prove that this is not the case. 

A first --and indirect-- proof is that such an additional velocity field would lead to an enhancement of mixing in all cases, while both an enhancement (salt-out) and a suppression (salt-in) are observed.

A second argument can be obtained from the experiment of T. S\'eon \textit{et al.} \cite{bib:Seonetal2007}, who studied the relative interpenetration of two fluids of different densities in a nearly horizontal configuration. 
While their flow takes place in a tube rather than a Hele-Shaw cell, they consider fluids of the same viscosity, just like in the present experiment. 
The Atwood number in their case, 
$\mathrm{At}=(\rho_2-\rho_1)/(\rho_2+\rho_1)$, where $\rho_1$ and $\rho_2$ are the densities of the fluids, ranges from $10^{-3}$ to $4\times10^{-2}$.  In our case $\mathrm{At}=\triangle\rho/(2\rho)=\beta\triangle S/2$, where $\rho$ is density, and $\beta$ is the expansion coefficient; with $\beta=2.4\, 10^{-2} \mathrm{M}^{-1}$ for LiCl 
\cite{bib:CRC_handbook_2011} 
and $\triangle S=20\, \mathrm{mM}$, we obtain $\mathrm{At}\sim 2.4\times10^{-4}$, which makes our configuration more stable from this point of view. 
In the particular case of a \textit{perfectly horizontal} tube, they obtain a \textit{decelerating} front, whose initial speed is based on the \textit{viscous} scales, that stops after some time. 
In our experiment, because of the vertical parabolic profile of the Hele-Shaw flow, such a velocity $v_\nu$ would scale like $\triangle\rho g\sim 12\mu v_\nu/h^2$ (where $g$ is gravity), \textit{i.e.} $v_\nu\sim\beta\triangle Sg h^2/(12\nu)\sim 6\,\mathrm{mm}.\mathrm{s}^{-1}$, superimposed to the pressure-driven basic flow. 
While the mean velocity of the front in the reference case is of order $2\,\mathrm{mm}.\mathrm{s}^{-1}$, this phenomenon (even if transitory) would lead to a velocity three times higher which would significantly change the positions of the fronts, between reference case and salt-in or salt-out;
however we did not observe any shift in the positions of the front between those three configurations.

The reason may be found in an article by L. Talon \textit{et al.} \cite{bib:Talonetal2013}:
In their computational paper, the flow takes place in a Hele-Shaw cell with a mean flow, like in our experiment, and fluids with different density and viscosity are considered. 
%They first obtain two-dimensional base states solutions of their problem. 
Because of gravity, they observe that the displacement front experiences a transitory state of higher velocity before reaching its stationary value; 
however, when the gravity parameter $F=\triangle\rho g h^2/(\mu U)$, which measures gravity versus viscous forces, is decreased towards unity, the transitory disappears. In our experiment this parameter, based on the velocity at the entrance of the chamber, is of order unity. 
Thus the flow-field in the three configurations (salt-in, reference and salt-out case) can be considered as identical, except for the density differences. 
\\
Past the early stages, the displacement front is stretched and folded by chaotic advection, which causes salt to begin to mix and stratify through a competition between gravity and diffusion:
a vertical gradient of salt appears, that can settle a colloidal movement because of diffusiophoresis.  
Following equation \ref{eq:vdp}, the vertical diffusiophoretic velocity $\mathbf{v}_{\mathrm{dp}}$ is of order $D_{\mathrm{dp}}\; \nabla S/S\sim D_{\mathrm{dp}}/h$. 
The typical time scale $\tau_{\mathrm{dp}}^{vert}$ associated to vertical diffusiophoresis is the time taken for a particle to go from half-depth where it is injected down to the bottom, hence $\tau_{\mathrm{dp}}^{vert}\sim h^2/(2D_{\mathrm{dp}})\sim8\,\mathrm{h}$. 
Although this is rather long, we could observe, when using a very thin laser sheet ($300\mu \mathrm{m}$-thick) that colloids tended to disappear below the sheet at large times in places of high stretching and folding rate.  
This is why we chose to illuminate the whole cell.

\end{document}